\documentclass[prb,twocolumn,nobibnotes,groupedaddress]{revtex4}
\usepackage{algorithm,algorithmic}
\usepackage{enumerate}
\usepackage{listings}
\usepackage{color}
\usepackage{float}
\usepackage{amssymb}
\usepackage{amsmath}

\usepackage{graphicx}
\graphicspath{{./Figures/}}

\definecolor{red}{rgb}{1,0.,0}

\begin{document}

%%\title{State Preparation Verification in Adaptive Variational Quantum Eigensolvers}
\title{Benchmarking Adaptive Variational Quantum Eigensolvers}
\date{\today}
\author{Daniel Claudino$^{1,2}$, Jerimiah Wright,$^{1,3}$ Alexander J. McCaskey,$^{1,2}$ and Travis S. Humble$^{1,3}$}
\affiliation{$^{1}$ Quantum Computing Institute,\ Oak\ Ridge\ National\ Laboratory,\ Oak\ Ridge,\ TN,\ 37831,\ USA \\
$^{2}$Computer Science and Mathematics,\ Oak\ Ridge\ National\ Laboratory,\ Oak\ Ridge,\ TN,\ 37831,\ USA \\
$^{3}$Computational Sciences and Engineering,\ Oak\ Ridge\ National\ Laboratory,\ Oak\ Ridge,\ TN,\ 37831,\ USA}

\begin{abstract}
By design, the variational quantum eigensolver (VQE) strives to recover the lowest-energy eigenvalue of a given Hamiltonian by preparing quantum states guided by the variational principle. In practice, the prepared quantum state is indirectly assessed by the value of the associated energy. Novel adaptive derivative-assembled pseudo-trotter (ADAPT) ansatz approaches and recent formal advances now establish a clear connection between the theory of quantum chemistry and the quantum state ansatz used to solve the electronic structure problem. Here we benchmark the accuracy of VQE and ADAPT-VQE to calculate the electronic ground states and potential energy curves for a few selected diatomic molecules, namely H$_2$, NaH, and KH. Using numerical simulation, we find both methods provide good estimates of the energy and ground state, but only ADAPT-VQE proves to be robust to particularities in optimization methods. Another relevant finding is that gradient-based optimization is overall more economical and delivers superior performance than analogous simulations carried out with gradient-free optimizers. The results also identify small errors in the prepared state fidelity which show an increasing trend with molecular size. 
\end{abstract}

%% requires nobibnotes 
\thanks{This manuscript has been authored by UT-Battelle, LLC, under Contract No.~DE-AC0500OR22725 with the U.S.~Department of Energy. The United States Government retains and the publisher, by accepting the article for publication, acknowledges that the United States Government retains a non-exclusive, paid-up, irrevocable, world-wide license to publish or reproduce the published form of this manuscript, or allow others to do so, for the United States Government purposes. The Department of Energy will provide public access to these results of federally sponsored research in accordance with the DOE Public Access Plan.}

\maketitle
%%%%%%%%%%%%%%%%%%%%%%%%%%%%%%%%%%%%%%%%
%%%%%%%%%%%%%%%%%%%%%%%%%%%%%%%%%%%%%%%%
\section{Introduction}
Quantum mechanics naturally lends itself to the description of phenomena at the atomic and molecular scale, including problems of chemical interest, which has culminated in the field of research known as quantum chemistry. Despite the formal impediments to achieve exact, closed-form solutions to quantum chemistry problems, there is a wide array of possible approximations, such as coupled cluster (CC) theory, \cite{shavitt_bartlett_2009} which have elevated quantum chemistry to good standing in the scientific community due to their reliability.
\par
In practice, CC faces two main difficulties that have hindered a more widespread adoption. One is that most of the success it has garnered over the years is due to its superior performance in the weak electron correlation regime, for which single-reference (SR) CC remains unchallenged. This success is justified because many problems in chemistry, such as thermochemistry, can be adequately treated as being largely weakly correlated. Yet, many other problems of interest involve molecules and materials that do not comply with this assumption, and for these instances, SR-CC breaks down. Despite multi-reference (MR) CC being an active area of research,\cite{monkhorst} theoretical and computational challenges currently curb the applicability of MR-CC.\cite{mrcc} 
\par 
A second obstacle to a more extensive use of CC theory is its pronounced computational cost. Reliable SR-CC methods, such as the so-called ``gold standard'' of quantum chemistry, coupled cluster singles and doubles (and perturbative triples), aka CCSD(T),\cite{ccsdt1, ccsdt2, ccsdt3} scale unfavorably with one-particle basis spanning the Hilbert space that houses the electronic wave function, which largely constrains the application of CCSD(T) to relatively small molecular systems. It is important to note that some of these limitations can be mitigated with methods such as configuration interaction (CI) in its MR formulation and the density matrix renormalization group (DMRG) which have in turn their own shortcomings, such as lack of size-extensivity and exactness contingent upon the dimensionality of the problem.
\par
Concurrent with developments in CC theory has been the increase in performance of computing technologies, which broadens the reach of computational chemistry techniques. Presently, this trend is continuing with the adaptation of chemistry methods, including CC, to the new technology paradigm of quantum computing. \cite{britt2017high,humble2019quantum} Because of the shared foundation in quantum mechanics, one of the most immediate applications for quantum computers is quantum chemistry.\cite{mcardle2020quantum} Recent advances have reformulated conventional problems in electronic structure for currently available quantum computing platforms.\cite{cao2019quantum} In particular, these efforts have led to a resurgence of the unitary coupled cluster (UCC) theory,\cite{romero2018strategies, ucc1, ucc2, ucc3} which can be employed in investigations where strong correlation is dominant. Quantum computing hardware appears to be well suited for building the states described by UCC, as this hardware can efficiently implement unitary operations to construct physical representations of the quantum state. {\color{black}Moreover, the intrinsic nature of the quantum computing logic can be exploited in order to propose new ansatze that, despite lacking a close connection to the underlying chemical intuition lent by UCC, are prone to a more efficient implementation, such as the so-called hardware efficient ansatz.\cite{kandala2017hardware}}
\par
%Borrowing from UCC, 
It is in the context of noisy intermediate-scale quantum (NISQ)\cite{Preskill2018quantumcomputingin} devices that the variational quantum eigensolver (VQE)\cite{vqe} has emerged as a promising method for testing the preparation and measurement of quantum states including those that represent the electronic eigenstates described by UCC.\cite{google2020hartree} Several variants of VQE are available,\cite{mc-vqe, mog-vqe} but all build on the variational principle from quantum mechanics, which constrains the quantum states that can satisfy the electronic eigenvalue problem.\cite{mcclean2016theory} While the initial VQE proposal assumes a predefined ansatz, this constraint has been relaxed, opening the door to adaptive approaches,\cite{adapt, iqcc} by which the preparable quantum states are driven by the problem at hand. In particular, the adaptive derivative-assembled pseudo-trotter (ADAPT) ansatz, which finds support on the recently coined ``disentangled'' UCC\cite{ucc_exact} and starts from an exact UCC representation of the electronic ground state to construct an approximate prepared state based on the dominant contributions. Early studies demonstrated this as a promising avenue for developing ansatze for specific molecules and constraints, such as highly accurate energetics or shallow circuits. 
\par
Here we benchmark adaptive VQE prescriptions, ADAPT-VQE in particular, by comparing the prepared quantum states with the conventional solutions obtained from exact diagonalization of the full configuration interaction Hamiltonian. We track the energy of the minimized expectation value as well as the fidelity of the corresponding prepared state using multiple ansatz, optimization methods, and molecular Hamiltonians. We calculate infidelity as a measure of error for the prepared quantum state relative to the expected, exact result from quantum chemistry using frozen-core Hamiltonians. Across these examples, we find that ADAPT-VQE is the more robust method due mainly to its performance with respect to optimization methods. While all methods lead to small errors as measured by the infidelity, these errors are found to grow with molecular size. 
\par 
This presentation is structured as follows. In Section \ref{sec:vqe}, we provided an overview of the ingredients in the VQE approach relevant to our purposes, followed by a short exposition of the underpinnings of ADAPT-VQE (Subsection \ref{ssec:adapt}) and a brief discussion on implementation of gradients and optimization in ADAPT-VQE (Subsection \ref{ssec:gradient}). The computational details permeating the reported simulations are exposed in Section \ref{sec:comp_details}. The main results are presented and discussed in Section \ref{sec:results} and several conclusions are drawn in Section \ref{sec:conclusion}.

\section{Variational Quantum Eigensolver}
\label{sec:vqe}
This section serves to illustrate the pertinent fundamentals of the VQE algorithm and to motivate the following exposition of adaptive ansatz construction. We start by recalling the variational principle, which is at the heart of VQE, and given as

\begin{equation}
    \label{eq:rayleigh}
    E \leq \min_{\Psi}\langle \Psi | \hat{H} | \Psi \rangle
\end{equation}
where $|\Psi \rangle$ is normalized {\color{black}trial wave function for which Equation \ref{eq:rayleigh} becomes an equality when $\Psi$ is constructed from a basis that spans the single-particle Hilbert space of all possible occupation numbers (the underlying Fock space)} and the electronic Hamiltonian $\hat{H}$ for a molecular system is given as

\begin{equation}
    \label{eq:hamiltonian}
    \hat{H} = \sum_{pq}h_{pq}p^\dagger q + \sum_{pqrs}h_{pqrs}p^\dagger q^\dagger sr
\end{equation}

The central problem in modern electronic structure theory is the description and quantification of the electron correlation from an un-entangled, mean-field wave function $|0\rangle$ whose preparation can be carried out in classical hardware in a timely fashion, e.g., Hartree-Fock (HF). In analogy with quantum chemistry, we can expect that there exists an operator that, once applied to $|0\rangle$, will account for the missing electron correlation. {\color{black}Bearing in mind that quantum computers manipulate quantum states in a well-defined Hilbert space, this configures a generic unitary operator $\hat{U}(\vec{\theta})$ whose main purpose is to build entanglement from an un-entangled reference function $|0\rangle$. The set of scalars $\vec{\theta}$ are parameters variationally} varied in order to minimize the expectation value in Equation \ref{eq:rayleigh}. With that, we recast Equation \ref{eq:rayleigh}:

\begin{equation}
    \label{eq:vqe}
    E \leq \min_{\vec{\theta}}\langle 0 |\hat{U}^\dagger (\vec{\theta}) \hat{H}_P \hat{U}(\vec{\theta})|0 \rangle
\end{equation}

In order to ensure that Equation \ref{eq:vqe} meets the requirements of quantum hardware, the fermionic, second-quantized operators found in the formulation of electronic structure problem, such as those in Equation \ref{eq:hamiltonian}, are brought to a qubit (spin) representation, with the additional constraint of fermionic anti-symmetry. Our approach uses the Jordan-Wigner transformation,\cite{JW} but others exist, and such a transformation yields $\hat{H}_P$ from $\hat{H}$, that is, the Hamiltonian in terms of strings of Pauli operators. Starting from the UCC ansatz, the unitary $\hat{U}(\vec{\theta})$ can be written as:

\begin{equation}
    \hat{U}(\vec{\theta}) = \text{exp}(\sum_k\theta_k(\hat{T}_k - \hat{T}^\dagger_k)) = \text{exp}(\sum_k\theta_k\tau_k)
\end{equation}
with the ${\hat{T}_k}$ representing the usual cluster operators in CC theory and $\tau_k = \hat{T}_k - \hat{T}^\dagger_k$, ensuring the anti-Hermiticity of the operators, which is a necessary condition for their utilization in quantum computing.

Once in possession of all ingredients in Equation \ref{eq:vqe}, the tasks of preparing the state $\hat{U}(\vec{\theta})|0 \rangle$ and measuring the terms in $\hat{H}_P$ are delegated to the quantum hardware, and $\hat{U}(\vec{\theta})$ is varied variationally with the aid of a classical optimization routine until $\langle \hat{H}_P \rangle$ {\color{black}reaches its minimum, which is dependent on the chosen optimizer and is taken as a good approximation to the sought} ground state energy. Due to the isomorphic property of the qubit mappings, $\langle \hat{H}_P \rangle = \langle \hat{H} \rangle$, yielding the lowest energy eigenvalue of the molecular Hamiltonian in Equation \ref{eq:hamiltonian}.

\subsection{ADAPT-VQE}
\label{ssec:adapt}

An important choice in the specification of the VQE method is the functional form of the ansatz $\hat{U}(\vec{\theta})$. Even for a relatively small Hilbert space, with a moderate number of cluster operators ${\hat{T}_i}$, the ansatz $\hat{U}(\vec{\theta})$ gives rise to a unitary that translates into multi-qubit gates and thus cannot be efficiently implemented in an actual quantum processor. Borrowing from the dynamics community, this can be alleviated by resorting to the Trotter-Suzuki decomposition, or Trotterization for short:

\begin{equation}
    \label{eq:trotter}
    \text{exp}(\sum_k\theta_k\tau_k) \approx \prod_k\text{exp}(\theta_k\tau_k)
\end{equation}
which here is limited to first-order.

The Adaptive Derivative-Assembled Pseudo-Trotter ansatz Variational  Quantum Eigensolver (ADAPT-VQE)\cite{adapt} takes advantage of Equation \ref{eq:trotter} to propose an iterative ansatz construction whereby only the perceived most relevant operator for energy lowering is added to the ansatz. {\color{black}A set of operators the algorithm can choose from needs to be provided, which in this work is comprised of the fermionic spin singlet adapted single and double excitations, borrowing from the usual UCCSD formulation, and subsequently mapped into the appropriate tensor products of Pauli operators via the Jordan-Wigner transformation. In principle, one could envision explicit enforcement or relaxation of other types of symmetry, and the effect of such choices on the performance of ADAPT is certainly a topic worth exploring. Moreover, the ADAPT algorithm has also recently been reported to perform well with other choices of operators, including a more economical pool of qubit operators,\cite{qubit_adapt} and has been applied to variational algorithms other than VQE.\cite{adapt_qaoa}} 

{\color{black}From a practical standpoint, at the $i$-th iteration of the algorithm, the energy gradient vector ($\mathcal{G}$) with respect to all $\{\theta_k\}$ in Equation \ref{eq:trotter} is computed from measurements on the circuit that prepares the state optimized in the previous iteration, represented by $|\psi_{i-1}\rangle$, with $|\psi_0\rangle = |0\rangle$. Labeling the energy at the current iteration $E_i$, we have:}

\begin{align}
    \label{eq:commutators}
    \mathcal{G} &= \left( \frac{\partial E_i}{\partial \theta_1}, \dots, \frac{\partial E_i}{\partial \theta_k}, \dots, \frac{\partial E_i}{\partial \theta_N} \right) \nonumber \\ &\frac{\partial E_i}{\partial \theta_k} = \langle \psi_{i-1}| [H, \tau_k] |\psi_{i-1}\rangle
\end{align}

{\color{black}and if the norm of this vector falls below a set threshold, the algorithm is deemed converged and the ansatz-growing loop is exited. Otherwise, the operator associated with the largest absolute component of $\mathcal{G}$ is selected to increment the ansatz:

\begin{equation}
    \label{eq:op_selection}
    |\psi_i\rangle = e^{\theta_i \tau_i} |\psi_{i-1}\rangle, \quad \tau_i = \{\tau_k | \max |\langle [H, \tau_k] \rangle_{i-1}|\}
\end{equation}

where $\langle [H, \tau_k] \rangle_{i-1}$ means this commutator was computed from observations in the circuit obtained from the previous iteration.}
With the selection of a new operator, the new ansatz is subject to the usual VQE routine and the corresponding energy mimimum is obtained. 

\subsection{Gradient Estimate and Classical Optimization in ADAPT-VQE}
\label{ssec:gradient}

From a quantum computing standpoint, ADAPT-VQE improves on VQE by potentially offering a more tractable circuit. However, this may come at the expense of a much larger number of measurements, as the evaluation of all $[H, \hat{A}_k]$ is performed at each iteration of the ADAPT loop{\color{black}, on top of the expected energy evaluations. In order to reduce the number of measurements associated with ADAPT-VQE simulations, adoption of a gradient estimate strategy can help improve the classical optimization step by reaching the sought minima with fewer calls to the hardware backend.}

To motivate the discussion, we start by invoking the gradient expression as introduced in the original formulation of ADAPT-VQE:

\begin{align}
    \label{eq:grad}
    \frac{\partial E}{\partial \theta_i} &= \langle \phi| \hat{H} \prod_{j=N}^{i+1}(e^{\theta_j\tau_j})\tau_i\prod_{k=i}^1(e^{\theta_k\tau_k}) |0\rangle \nonumber \\
    & - \langle 0| \prod_{k=1}^i(e^{-\theta_k\tau_k})\tau_i\prod_{j=i+i}^N(- e^{\theta_j\tau_j})\hat{H}|\phi\rangle
\end{align}
where {\color{black}$\prod_ie^{\theta_i\tau_i}|0\rangle = |\phi\rangle$}.

Equation \ref{eq:grad} can be further simplified into a recursive formula:

\begin{equation}
    \label{eq:grad_recursion}
    \frac{\partial E}{\partial \theta_i} = \langle \phi| \left[\hat{H},  \prod_{j=N}^{i+1}(e^{\theta_j\tau_j}) \tau_i \prod_{j=i+1}^N(e^{-\theta_j\tau_j}) \right]|\phi \rangle
\end{equation}

{\color{black}Before moving further in the discussion regarding the use of gradients to support the classical optimizer, let us clarify a potential source of confusion. At a certain ADAPT-VQE iteration, the circuit previously optimized is implemented to prepare the state from which the current iteration builds upon. The gradient vector $\mathcal{G}$ is then computed upon the necessary measurements for all $\tau_k$ in the chosen operator pool (Equation \ref{eq:commutators}), and the operator that has the largest commutator (in absolute value) is selected. And this is the extent to which the gradient is employed at this stage. On the other hand, we now have a new ansatz, which is composed of all previously added operators that enable preparation of $|\psi_{i-1}\rangle$, along with the newly added operator from Equation \ref{eq:op_selection}. Each of these operators have a corresponding variational parameter, which in the following VQE step need to be re-optimized. It is in this optimization that we would employ the gradients as written in Equations \ref{eq:grad} and \ref{eq:grad_recursion}, and whose magnitude needs to be minimized in order to signal the finding of an extremum (minimum in this case). For an operator pool containing $N$ elements, at each ADAPT-VQE iteration, all $N$ elements of $\mathcal{G}$ need to be evaluated, but the magnitude of this vector is not directly minimized by varying the circuit parameters, only indirectly by the addition of enough operators in the ansatz. On the other hand, for optimization purposes, at the $i$-th iteration, only $i$ gradient elements are considered, and the search for the energy minimum is guided by the minimization of the magnitude of this $i$-th dimensional gradient vector. Finally, another crucial point worth pointing out is that the commutators in Equation \ref{eq:commutators} are equivalent to those in Equation \ref{eq:grad_recursion} only for the operator most recently added, i.e., $\tau_i$ in Equation \ref{eq:op_selection}.}

For the purposes of an economical quantum resource utilization, it is desirable to deploy only one circuit to be used in both energy and gradient estimates (the same circuit is implemented many times, one for each term in the Hamiltonian/gradient). Even though the recursive formula in Equation \ref{eq:grad_recursion} could, in principle, satisfy this requirement, this commutator cannot be measured.\cite{circuit_learning} As originally proposed, the gradient is no longer given in an expectation value form, requiring an auxiliary state to be prepared via introduction of ancilla qubits, which deviates from our requirement of saving quantum resources. For that reason, we resort to numerical finite differences as means of carrying out gradient-based optimizations in the current work. 

In terms of resource estimation, for a circuit depth of $\mathcal{O}(N)$, forward or backward finite differences are akin to introducing a single $R_z(h)$, where $h$ is the step size, leading the a circuit depth of $\mathcal{O}(N+1)$, while the use of central differences, thus, has circuit depth of $\mathcal{O}(2(N+1))$, the former being used here due to its superior convergence properties. This is the cost incurred in the numerical gradient estimate for each parameter being optimized and a detailed discussion is provided in Subsection \ref{ssec:resource}. Such an estimate may be improved with strategies such as the quantum natural gradient\cite{qng} or exploiting partial tomography.\cite{jacobi} These ramifications are worthy of a separate study, and will not be further investigated here.

\section{Computational Details}
\label{sec:comp_details}

The quantum simulations detailed in this manuscript were performed using the VQE and ADAPT-VQE algorithms and numerical gradient strategies as implemented in the XACC hybrid quantum-classical computing framework,\cite{xacc1, xacc2} with the latter algorithm leveraging a convergence criterion of $||\mathcal{G}|| \leq 10^{-2}$. {\color{black}We emphasize that this parameter can be of substantial impact on the results, as it controls the size of the obtained ansatz. In light of the findings in Ref. \citenum{adapt}, the adopted value in this paper is believed to strike a satisfactory balance between accuracy and circuit depth}. The resulting circuits were simulated via the TNQVM (tensor-network quantum virtual machine)\cite{tnqvm} XACC simulation backend and employed a noiseless, matrix product state (MPS) wave function decomposition for the quantum circuit with the aid of the ITensor library.\cite{itensor} {\color{black}XACC provides other simulation backends, as well as physical backends targeting QPUs from IBM and Rigetti. For the size of the problems studied in this work, there may not be perceived benefits from choosing TNQVM over other XACC simulation backends like Aer\cite{Qiskit} or QPP.\cite{qpp} TNQVM is expected to be advantageous over other simulation approaches for problems requiring more qubits,\cite{tnqvm} but we leave this to future work and do not investigate it here.}

The COBYLA\cite{cobyla} algorithm was used as a gradient-free optimizer, while gradient-based optimizations were carried out with the L-BFGS algorithm,\cite{lbfgs1, lbfgs2} {\color{black}with all parameters being initialized at 0 at each optimization cycle for both optimizers. Other approaches have been reported in the literature, such as random initialization,\cite{trotter_uccsd} or as in the original implementation of ADAPT-VQE,\cite{adapt} where the new parameter is initialized at 0, while the previous parameters are initialized from the optimal values obtained in the previous ADAPT iteration.} XACC offers both optimizers via its interface with \texttt{NLOpt}.\cite{nlopt}

The potential energy curves (PEC) of NaH and KH, were generated by imposing the frozen-core approximation, reducing the number of configurations to only those arising from one $\sigma$ orbital and its $\sigma^*$ counterpart, that is, a two electrons in two orbitals complete active space (CAS(2,2)) problem. The one- and two-electron integrals necessary for the construction of the Hamiltonians and the corresponding references CAS energies were obtained with PySCF,\cite{pyscf} with all calculations performed with the STO-3G basis set.\cite{sto3g1, sto3g2, sto3g3}

The quality of the output circuits in preparing the desired states is assessed via the fidelities computed with respect to the ground state full configuration interaction (FCI) wave function. This corresponds to the lowest energy eigenvector from exact diagonalization in the $2^N$ Hilbert space, with orbital occupation determined by the number of electrons. In possession of the circuits from the quantum simulations, the respective state vector representation is obtained using the XACC interface to the Qiskit Aer simulator.\cite{Qiskit}

\section{Results and Discussion}
\label{sec:results}

Typically, the quality of the state obtained from the variational optimization of the gate parameters is probed indirectly by comparison of the computed energies with trustworthy references values or the exact lowest energy eigenvalue whenever computationally feasible. Thus, we start by investigating the energy profile along the atomic displacement, and subsequently contrast these findings with the analysis of the appropriateness of the corresponding states via evaluated fidelities relative to the vector corresponding to the lowest eigenvalue in the active space.

\subsection{Potential Energies Curves}
\label{ssec:pec}

\begin{figure}[ht!]
    \centering
    \includegraphics[width=\columnwidth]{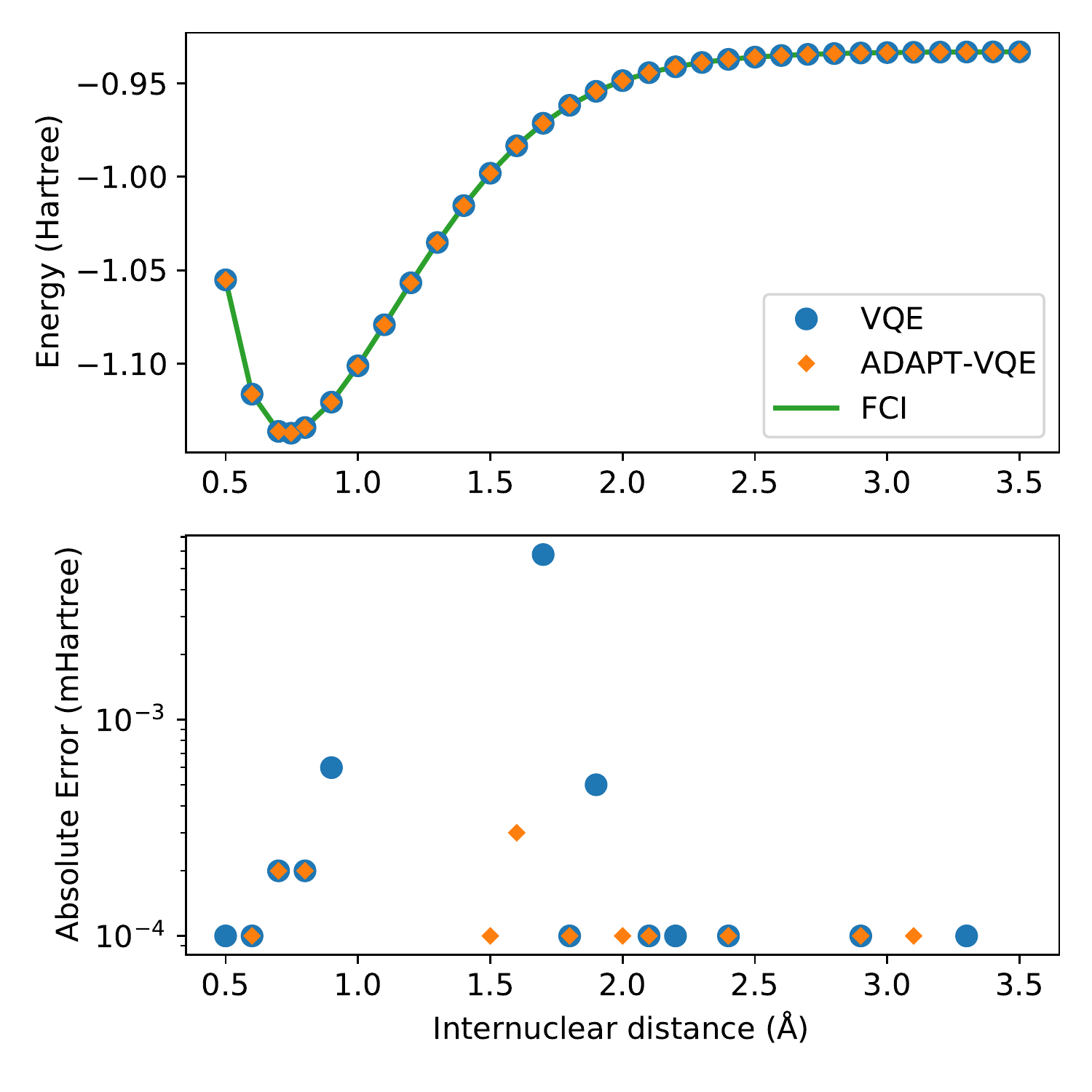}
    \caption{(top) Potential energy curves of H$_2$ computed with the STO-3G basis set for FCI (green solid line), VQE (blue circles), and ADAPT-VQE (orange diamonds) with the COBYLA optimizer. (bottom) Absolute error in the minimized energy for VQE (blue) and ADAPT-VQE (orange) relative to the FCI reference value.} 
    \label{fig:h2_energies}
\end{figure}

We start investigating the behavior of the energy by studying the H$_2$ molecule. This example has been extensively approached in quantum computing, and hardly poses any difficulty, at least from the standpoint of numerical simulations, as opposed to deployment to actual hardware. However, it serves as a baseline for the following discussion, as the orbital spaces in the other molecules are reduced to an active space with the goal of resembling the H$_2$ molecule. Results with the VQE and ADAPT-VQE ansatze are plotted in Figure \ref{fig:h2_energies}, along with FCI results.

Unsurprisingly, there is a remarkable agreement between simulated and exact values, both qualitatively and quantitatively. Absolute errors from FCI are found in the sub-miliHartree range throughout the energy scan, and with either choice of ansatz, the observed errors would be inconsequential when taking into account the scale of the errors introduced by noise in the operation of quantum devices. %Despite small, the errors incurred by the ADAPT-VQE are smaller for a larger number of interatomic distances.
The impression that some points %referring to ADAPT-VQE 
are ``missing'' from the bottom plot of Figure \ref{fig:h2_energies} is explained by these values being numerically identical to the FCI values (to seven decimal places), hence not being plotted in the logarithmic scale.

\begin{figure}[ht!]
    \centering
    \includegraphics[width=\columnwidth]{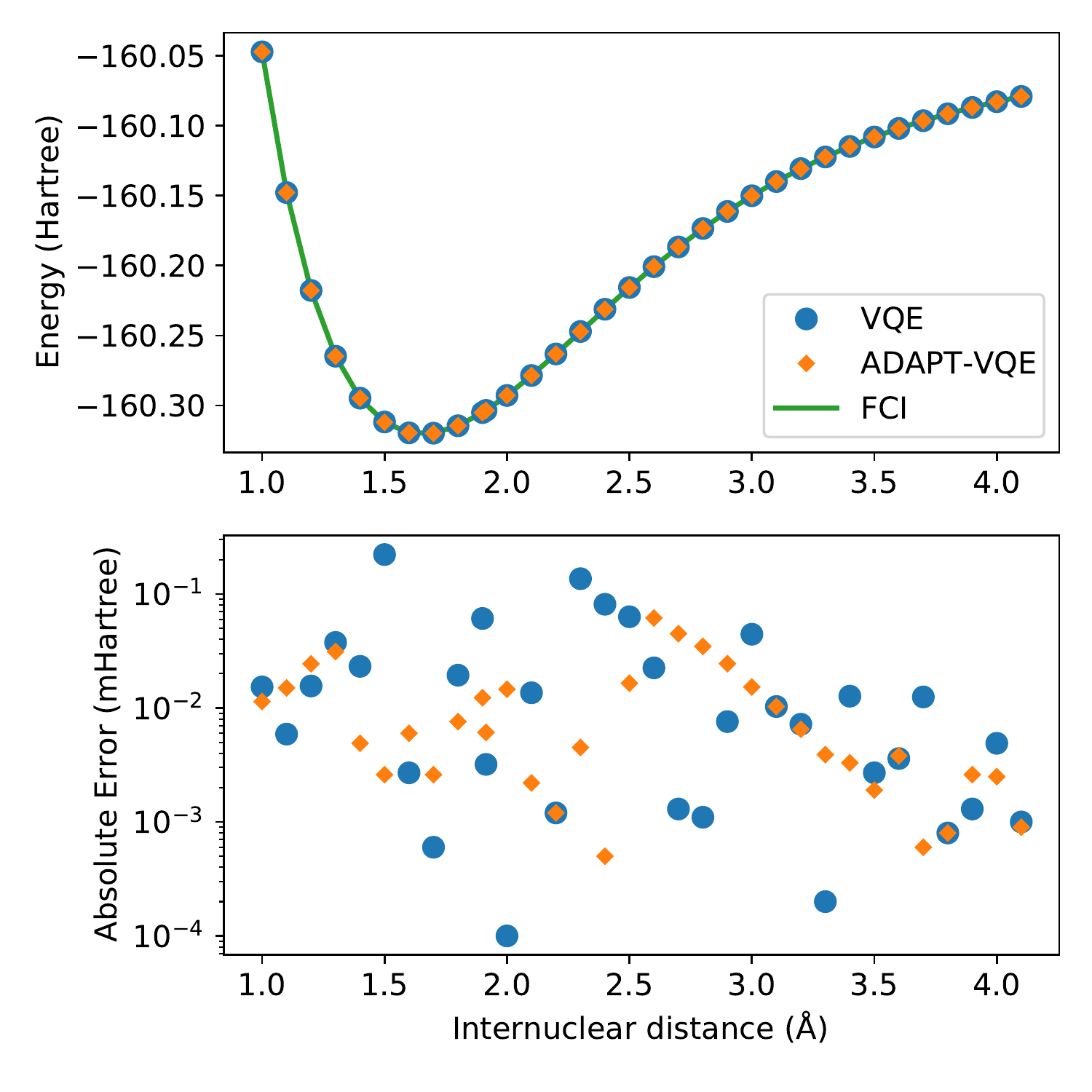}
    \caption{(top) Potential energy curves of NaH computed with the STO-3G basis set for FCI (green solid line), VQE (blue circles), and ADAPT-VQE (orange diamonds) with the COBYLA optimizer. (bottom) Absolute error in the minimized energy for VQE (blue) and ADAPT-VQE (orange) relative to the FCI reference value.}
    \label{fig:nah_energies}
\end{figure}

The results from the potential energy curve from simulations on the NaH molecule are presented in Figure \ref{fig:nah_energies}.

Visual inspection of the top plot reveals that the choice between the two ansatze being considered here yield energies that track one another very well, but because of the energy scale of this plot, it begs a closer look. The bottom plot displays the absolute errors between VQE and ADAPT-VQE with respect to FCI. The errors here are still within chemical accuracy ($<$1 kcal/mol), and are unlikely to be of much relevance in the total error if such simulations are executed in a quantum computer. However, there is a clear trend of increase in the magnitude of the computed deviations when compared to the hydrogen molecule, whose results are in Figure \ref{fig:h2_energies}.

\begin{figure}[ht]
    \centering
    \includegraphics[width=\columnwidth]{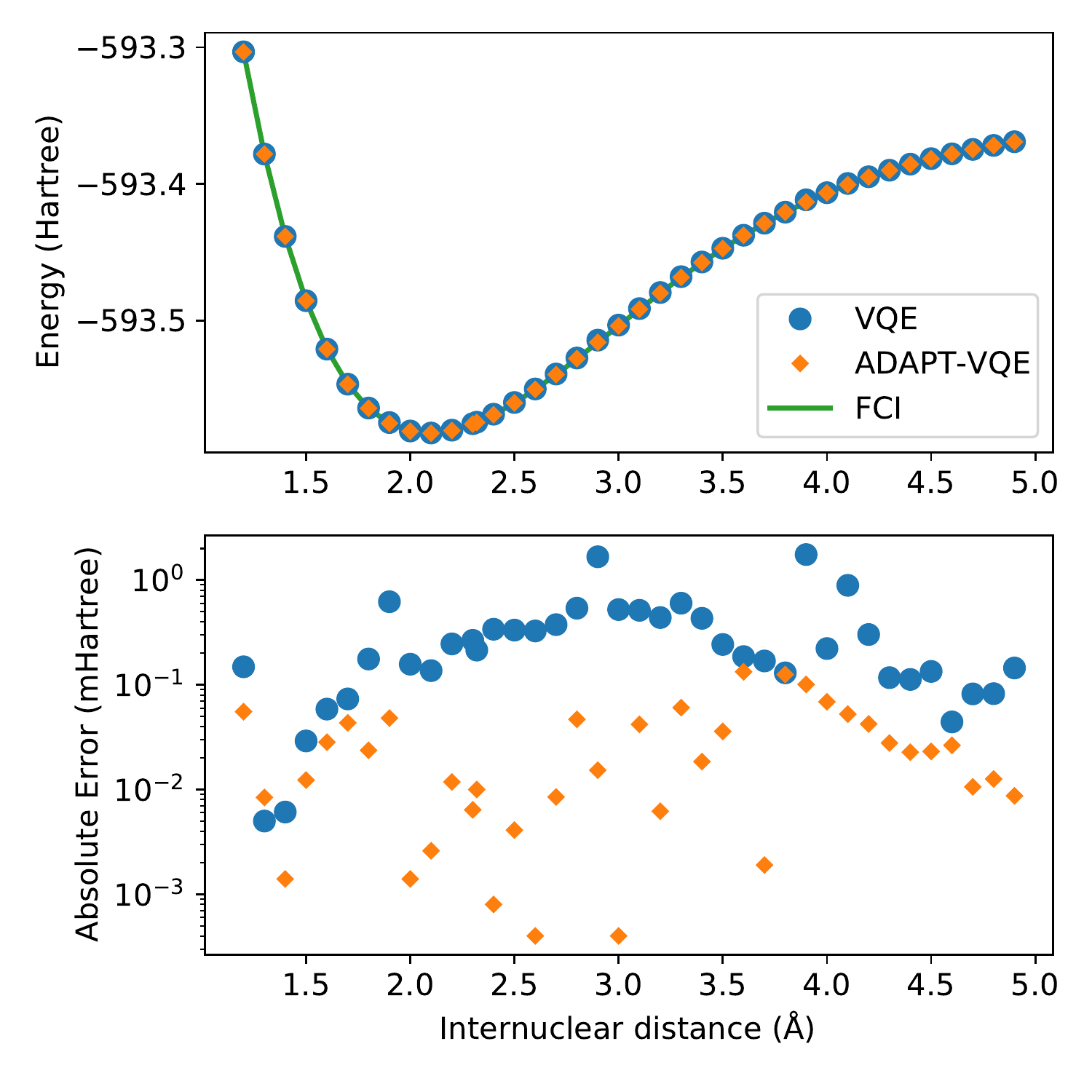}
    \caption{(top) Potential energy curves of KH computed with the STO-3G basis set for FCI (green solid line), VQE (blue circles), and ADAPT-VQE (orange diamonds) with the COBYLA optimizer. (bottom) Absolute error in the minimized energy for VQE (blue) and ADAPT-VQE (orange) relative to the FCI reference value.}
    \label{fig:kh_energies}
\end{figure}

In Figure \ref{fig:kh_energies} we again observe some of the patterns that follow from the analysis of Figures \ref{fig:h2_energies} and \ref{fig:nah_energies}. The energy scale here is much too large to able to reveal relatively minor inadequacies, even though qualitative discrepancies, such as those arising from symmetry breaking or the crossing of lines of different states, would be evident had they been present. The bottom plot, exhibiting the energy differences from FCI, offers a more reliable evidence, allowing us to infer that ADAPT-VQE is overall superior, with smaller errors for the vast majority of points (the exception being 1.4 \AA). Perhaps more importantly, we observe a general trend of the points from simulations with the plain VQE ansatz approaching the 1 mHartree, with the distances of 2.9 and 3.9 {\AA} now found more than 1 kcal/mol above the respective FCI energy. 

\subsection{Optimization Strategies}
\label{ssec:optimization}

The potential energy curves presented and discussed in Subsection \ref{ssec:pec} are based upon gradient-free optimization carried out with the COBYLA optimizer. We report that analogous simulations were performed with the Nelder-Mead optimizer, which is also a gradient-free alternative, but preliminary investigations pointed to COBYLA being a superior choice, at least for the chosen molecules. To contrast the performance of gradient-free optimization in the current context, we use the L-BFGS optimizer for parameter update, as implemented in \texttt{NLOpt}, with gradient estimated via central numerical finite differences. To assess the relative performance of these two approaches as the bond in the current diatomic molecules is stretched, we plot the difference between energies obtained with the COBYLA optimizer and those with L-BFGS+finite differences, that is, $E\text{(COBYLA)}-E\text{(L-BFGS)}$. That way, positive energy differences indicate there is an improvement by turning to a gradient-based optimization, while the opposite signals that the current gradient-free method reached a lower energy. 

\begin{figure}[ht!]
    \centering
    \includegraphics[width=\columnwidth]{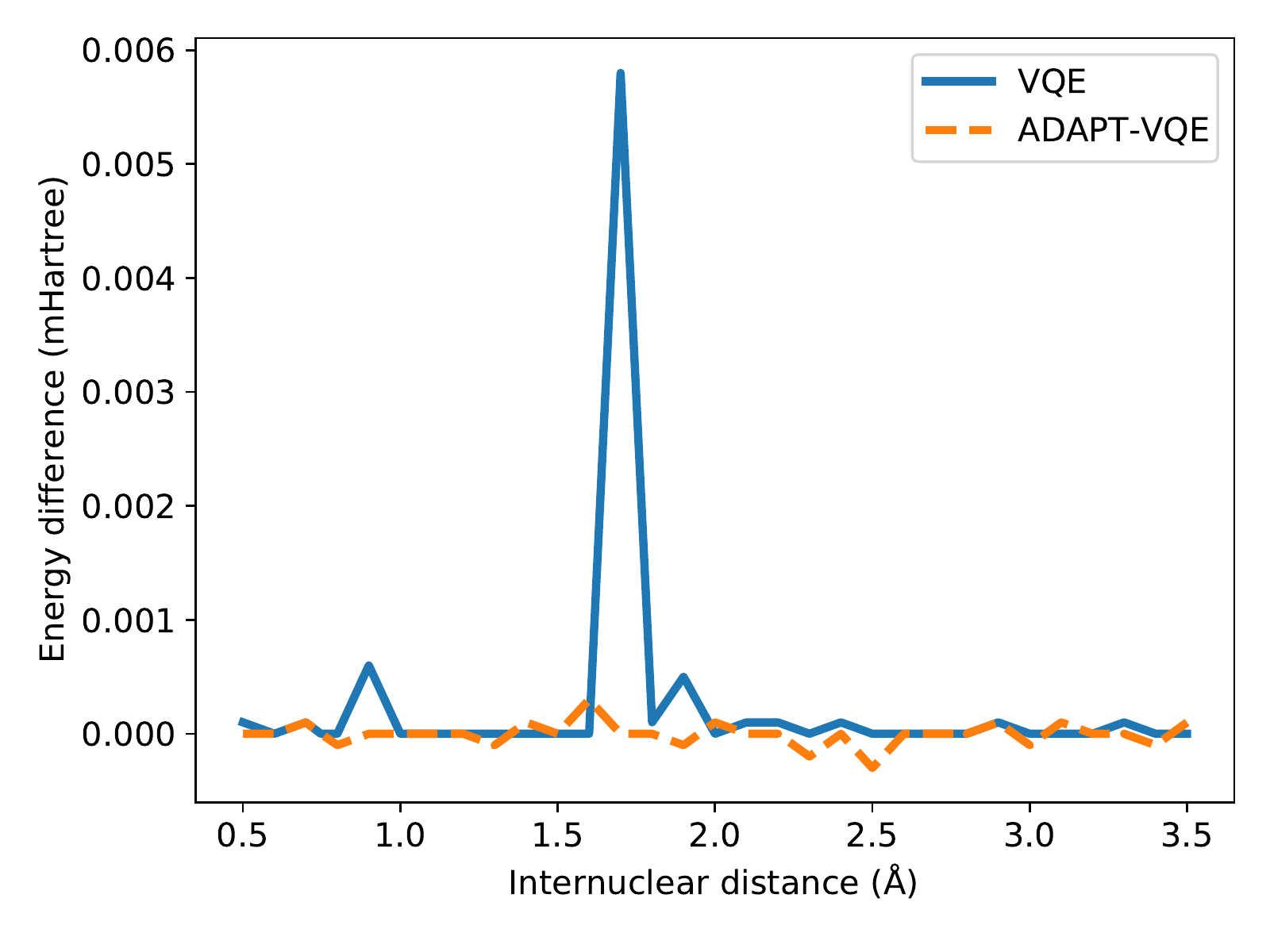}
    \caption{Difference between the energies from COBYLA and L-BFGS optimization with central finite differences for the H$_2$ potential energy curve.}
    \label{fig:h2_gradients}
\end{figure}

We observe compatible energies for the H$_2$ case, regardless of the underlying optimization strategy, for the entirety of Figure \ref{fig:h2_energies}. In order to maintain consistency, we plot the energy difference between the two optimization prescriptions in a miliHartree scale, and the spike in $E\text{(COBYLA)}-E\text{(L-BFGS)}$ in 1.7{\AA}, when rationalized with the scale in mind, shows a deviation in the $\mu$Hartree range. Due to the presence of all the many-body operators necessary for exactness,\cite{ucc_exact} we expect and in fact observe results on par with the numerical precision imposed by the employed optimizers ($10^{-6}$ Hartree in relative energy).
%, until we cross 2.5{\AA}. There are two clear discrepancies in 2.7 and 3.0 {\AA} when the energy minimum is approached with the COBYLA optimizer, in line with Figure \ref{fig:h2_energies}. These peaks are largely mitigated by turning to gradient-based alternative, which alludes to a more robust search for the optimal set of parameters when guided by an estimate of the corresponding gradients.

\begin{figure}[ht!]
    \centering
    \includegraphics[width=\columnwidth]{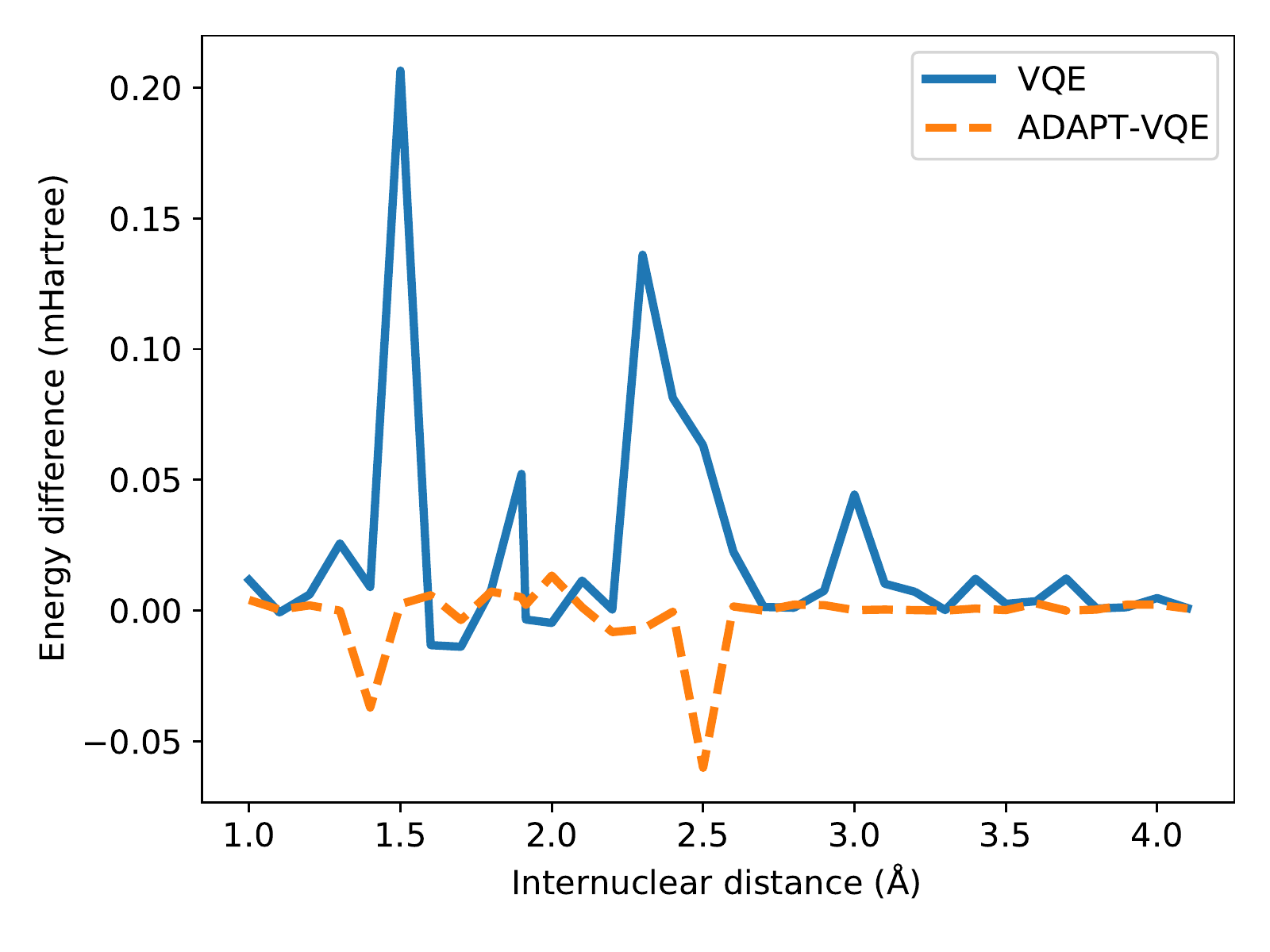}
    \caption{Difference between the energies from COBYLA and L-BFGS optimization with central finite differences for the NaH potential energy curve. }
    \label{fig:nah_gradients}
\end{figure}

While most of the PEC for H$_2$ showed no major dependence on the adopted optimization procedure, the picture is significantly different in the case of NaH, as portrayed in Figure \ref{fig:nah_gradients}. Even though the values for $E\text{(COBYLA)}-E\text{(L-BFGS)}$ are still rather small, in the sub-miliHartree range, noticeable differences are more frequent here. Albeit of $\mu$Hartree in magnitude, we also observe cases where COBYLA provides a lower energy than L-BFGS, most notably for ADAPT-VQE in the 1.4 and 2.5 {\AA} interatomic distances. On the other hand, in an overall assessment of the performance between VQE and ADAPT-VQE, the latter displays a more pronounced insensitivity with respect to the choice of optimization scheme.

\begin{figure}[ht!]
    \centering
    \includegraphics[width=\columnwidth]{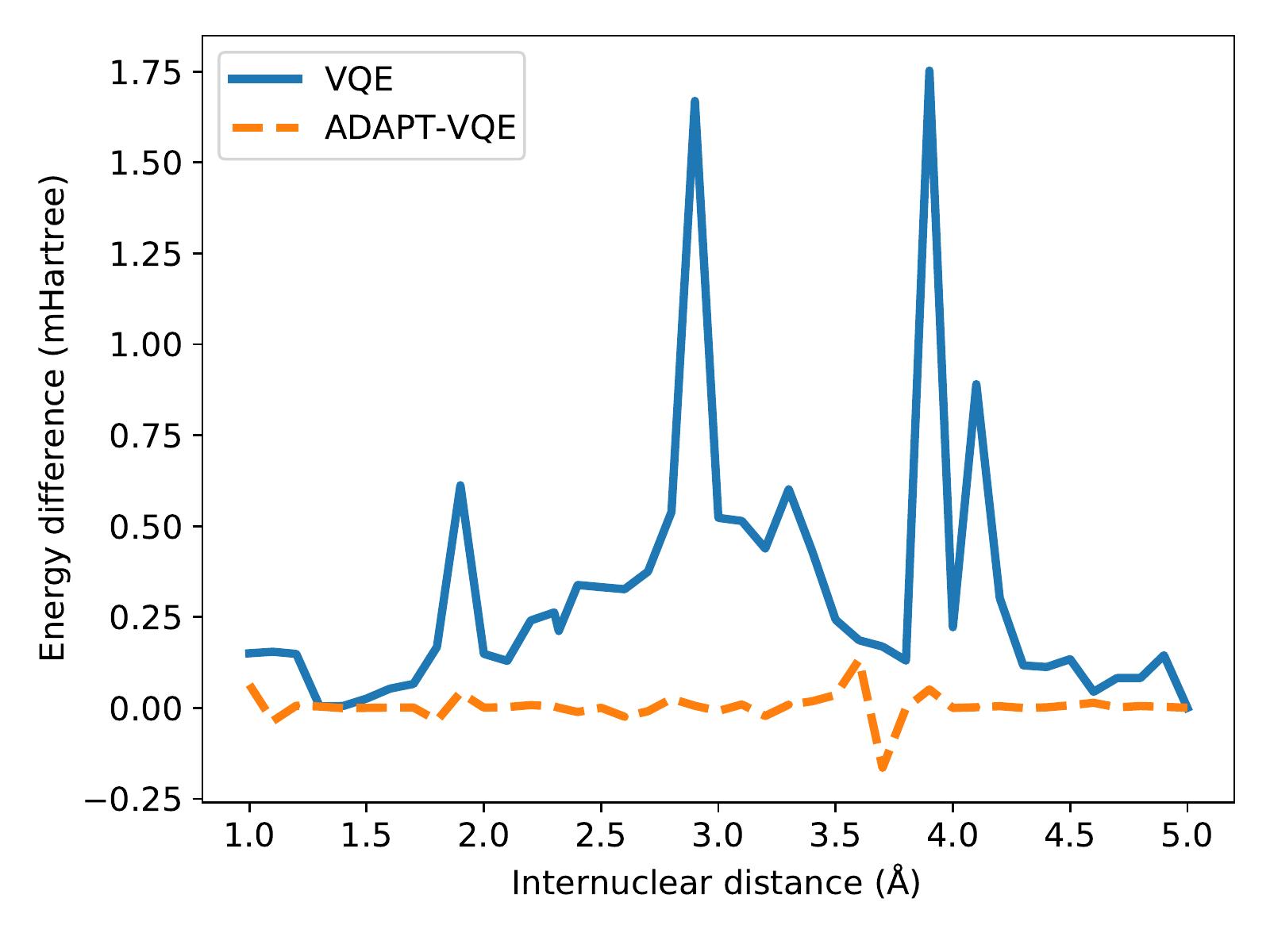}
    \caption{Difference between the energies from COBYLA and L-BFGS optimization with central finite differences for the KH potential energy curve.}
    \label{fig:kh_gradients}
\end{figure}

An even more drastic contrast is found from inspection of Figure \ref{fig:kh_gradients}, where $E\text{(COBYLA)}-E\text{(L-BFGS)}$ are plotted for the KH molecule. Some of the qualitative assertions pointed out in Figure \ref{fig:nah_gradients} hold, namely that the performance of VQE is much more influenced by the choice of optimization strategy than ADAPT-VQE. Not only that, but ADAPT-VQE is largely unaffected by employed optimizer, at least between the two alternatives in consideration. Here again, the differences seen for VQE correlated well with the deviations from FCI reported in Figure \ref{fig:kh_energies}, further corroborating the claim that a gradient-based optimization, given the current conditions, is a more robust for approaching the lowest energy eigenvalue of molecular Hamiltonians.

\subsection{State Fidelities}
\label{ssec:fidelities}

As previously stated, energy values can be used as valuable metric of the adequacy of a given set of variational parameters and trial state. However, the energy alone may not be indicative of the quality of the corresponding state and even acceptable energy values do not guarantee equally satisfactory values for other properties. The usual electronic Hamiltonian, as shown in Equation \ref{eq:hamiltonian}, transforms as the most symmetric irreducible representation for a given point group, therefore yielding the same energy in the case of degenerate states. Other operators, however, such as the terms in the multipole expansion of the electric potential, do not display this feature, meaning that degenerate states may yield different expectation values for such operators.

In order to examine the state prepared by the two circuit approaches considered here, we compute their ``infidelities'' with respect to the exact FCI state within the aforementioned active spaces, which is mathematically represented by $1-|\langle \Psi_\text{FCI}|\hat{U}(\vec{\theta})|0 \rangle|$, where $\vec{\theta}$ here are the set of optimal values also utilized for the energy computations in Subsections \ref{ssec:pec} and \ref{ssec:optimization}. We acknowledge that, while this provides direct inroads in the state being output at completion of the state preparation, it cannot be experimentally realized. However, in the case of moderately sized molecules for which the exact diagonalization of the Hamiltonian is feasible, this can provide valuable insights.

The energy differences discussed in the case of the hydrogen molecule in Subsection \ref{ssec:pec} and \ref{ssec:optimization} are quite small when considering the magnitude of the other potential sources of error that can arise in the presence of noise, either through a model or in the operation of an actual quantum device. Due to the simplicity of the electronic structure of this molecule the state prepared according to the two ansatze construction prescriptions investigated here yield infidelities that are below the numerical thresholds employed here, and certainly would be unnoticeable for realistic purposes. Because they offer little insight, we abstain from plotting the infidelity results for this molecule here.

\begin{figure}[ht!]
    \centering
    \includegraphics[width=\columnwidth]{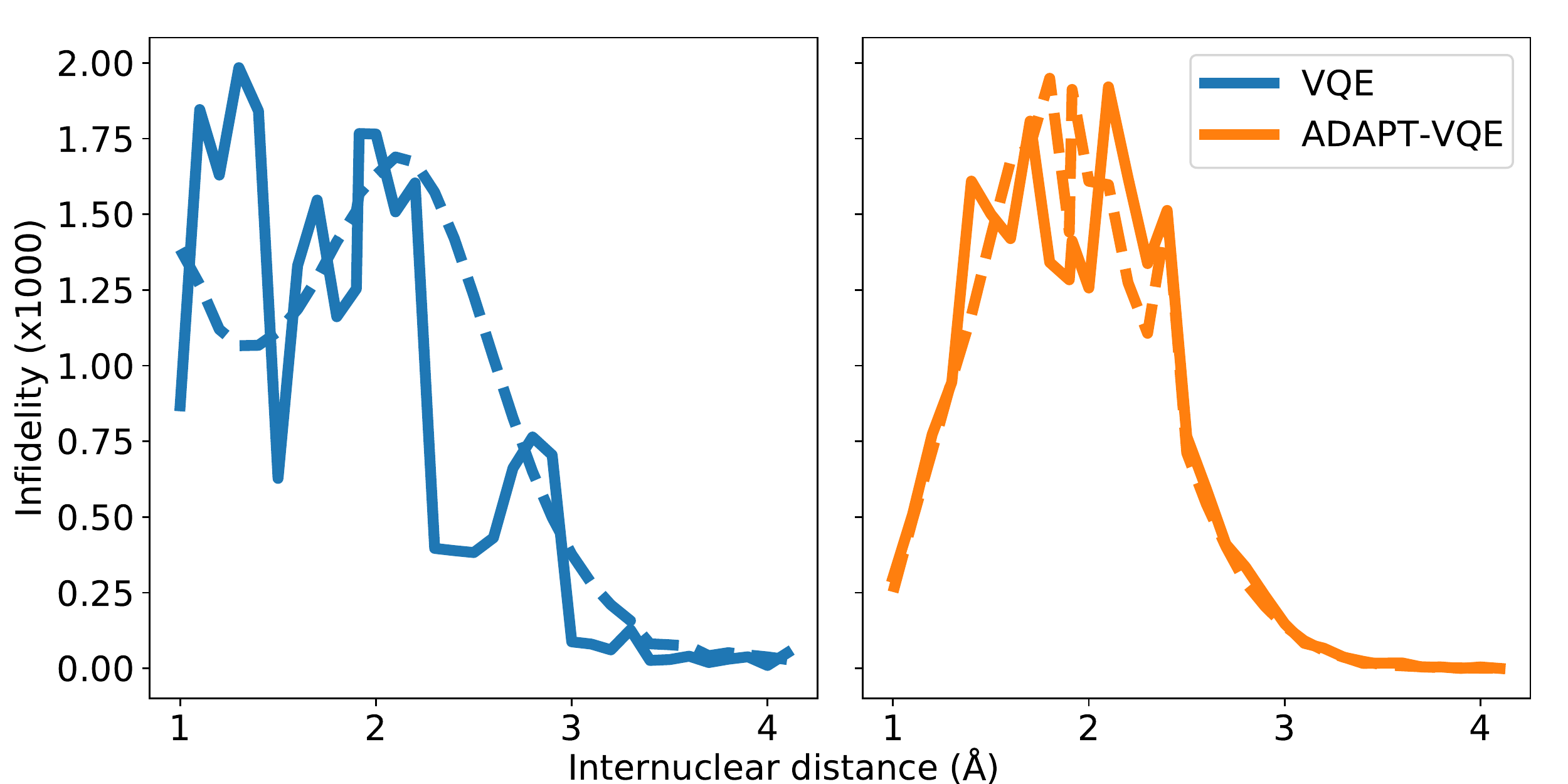}
    \caption{State infidelities for VQE and ADAPT-VQE using COBYLA (solid line) and L-BFGS (dashed line) optimization with central finite differences for the NaH potential energy curve.}
    \label{fig:nah_fidelity}
\end{figure}

Before delving into the particularities of each curve in Figure \ref{fig:nah_fidelity}, we bring the reader's attention to the scale of the plots, signaling a remarkable agreement between the state prepared and the one expected (FCI). It should come as no surprise that the largest infidelities are found in the vicinity of the Coulson-Fischer point, the most demanding region in the energy landscape, and subsequently approach zero as the atoms are moved far apart. The infidelities for the VQE ansatz follow a smooth progression when employed in conjunction with the gradient-based optimizer L-BFGS, whereas the same is not true for the other combinations of ansatze and optimization. This is likely a compound effect, explained by the former being a fixed circuit, where only the associate $\vec{\theta}$ changes throughout the energy scan. The latter, however, can assume a different composition, changing according to the demands of the electronic structure at each bond length. This works along with the fact that gradient-based optimization, at least in the current study, provides a tighter, more reliable solution. {\color{black}For the NaH and KH cases, we plot the number of operators in the ansatz in Figure \ref{fig:operators}.

\begin{figure}[ht!]
    \centering
    \includegraphics[width=\columnwidth]{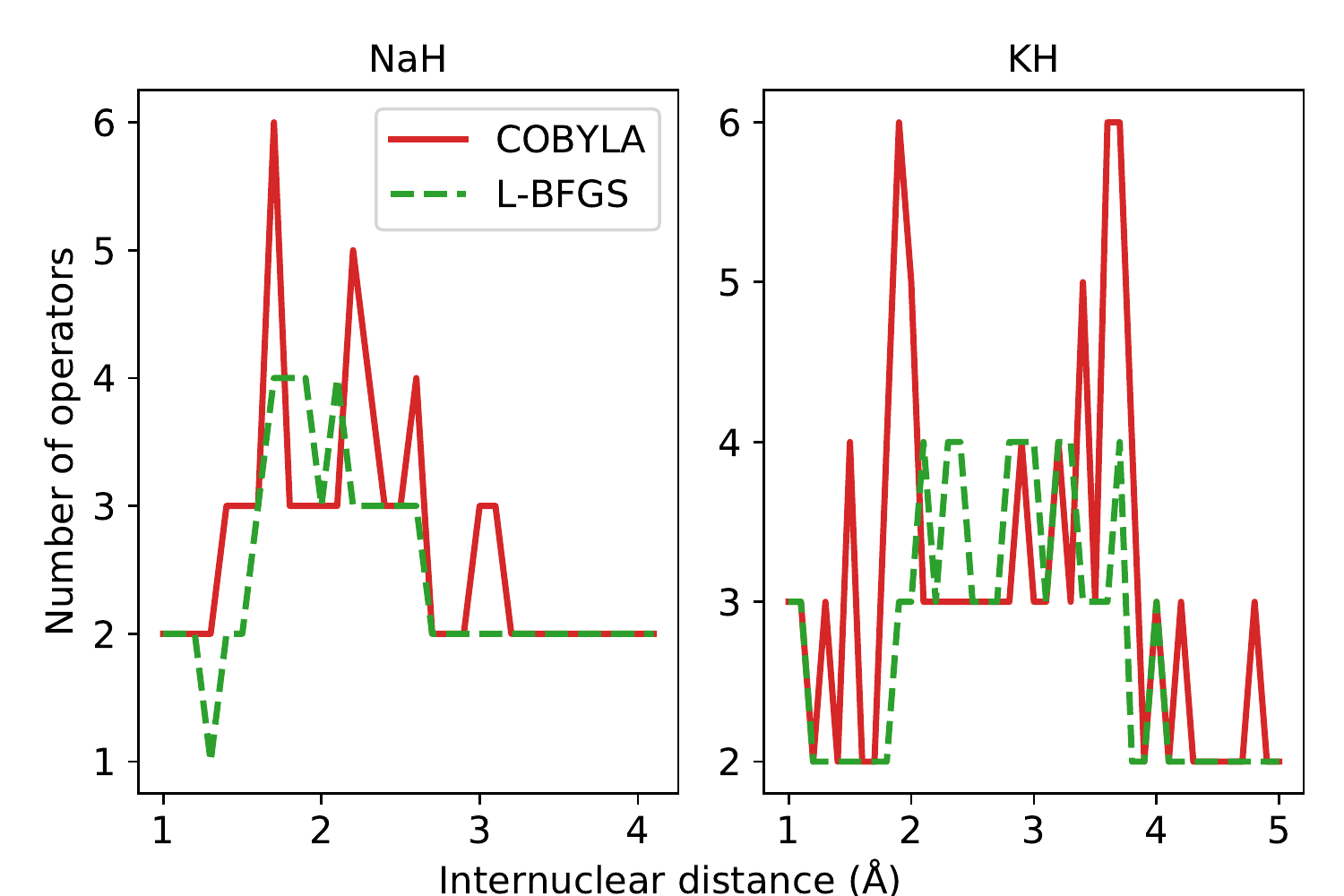}
    \caption{Number of operators in the ADAPT-VQE ansatze using the COBYLA (solid line) and L-BFGS (dashed line) optimizers. The corresponding VQE ansatz has two operators.}
    \label{fig:operators}
\end{figure}

Once again, there is a clear advantage in turning to gradient-based optimization, as it renders ansatze with fewer operators. For some internuclear distances, the ADAPT-VQE ansatze, even when optimized with L-BFGS, contain more operators than the corresponding VQE ansatz. This is not necessarily in contradiction with the some of the findings from Ref. \citenum{adapt} because those results were obtained for different molecules and using different optimization implementations. Yet, we would expect that when comparing against a larger VQE problem, such as those investigated in that paper, we would see similar trends.}
We also speculate that another variable that can contribute to the observed behavior is the tolerance that controls how tight the optimization should be. Because we are using a default $10^{-6}$ threshold in relative energy as the tolerance and there is no clear connection between the quality of the energies and the respective prepared states, the absolute energies values may fall in a scale that may have a small, but non-negligible effect on the fidelities, which is also evidence of the effect it can have in the output state, further corroborated by the number of operators found in the respective ansatze, yet not enough to alter any of the main conclusions drawn from the results presented here.

\begin{figure}[ht!]
    \centering
    \includegraphics[width=\columnwidth]{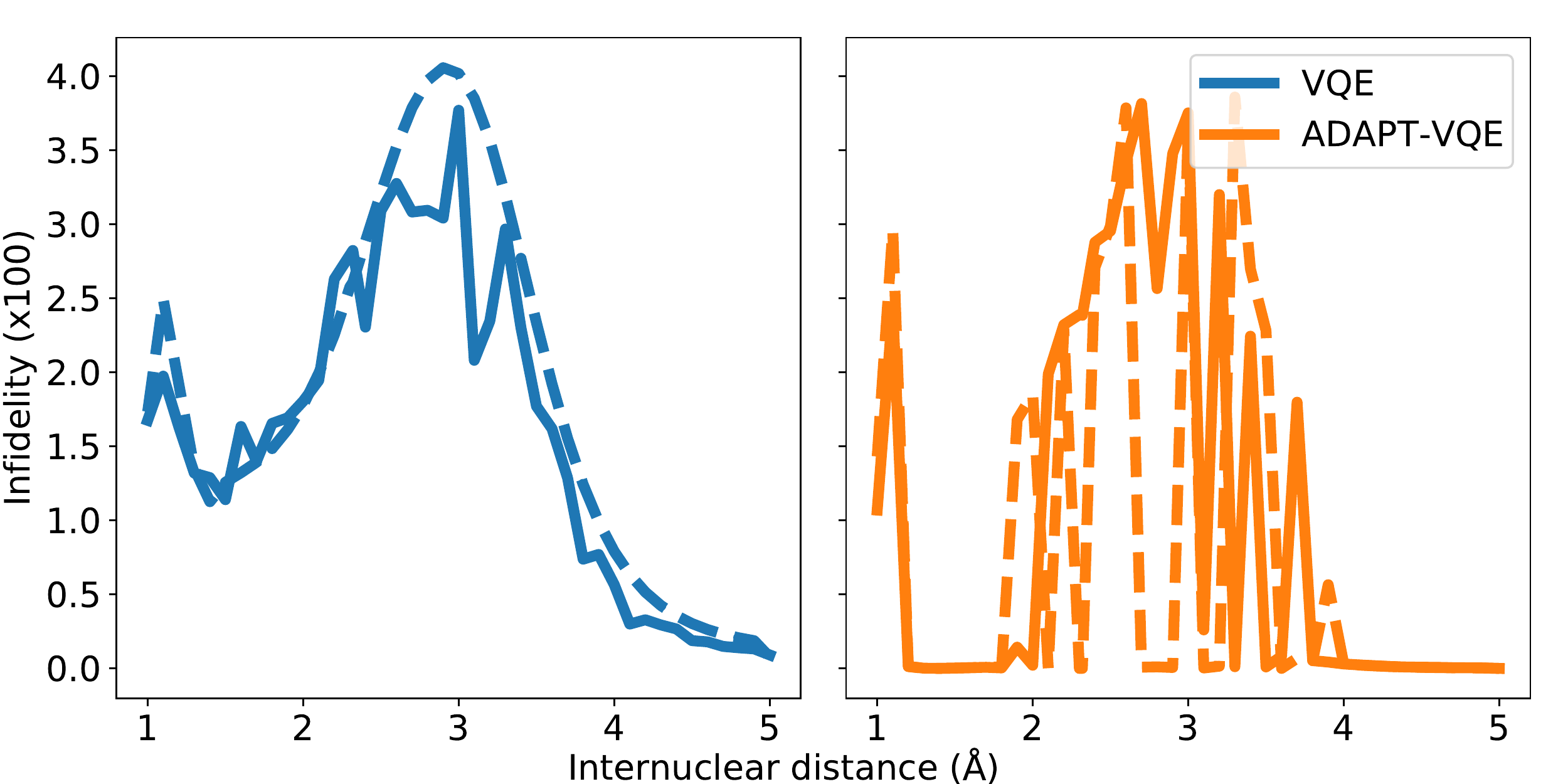}
    \caption{State infidelities for VQE and ADAPT-VQE using COBYLA (solid line) and L-BFGS (dashed line) optimization with central finite differences for the KH potential energy curve.}
    \label{fig:kh_fidelity}
\end{figure}

Many of the main inferences from the analysis of the Figure \ref{fig:nah_fidelity} hold for the KH molecule, whose infidelities are shown in Figure \ref{fig:kh_fidelity}. Firstly, the infidelities, though still quite small, are about an order of magnitude larger. The smoothness and overall profile observed for the VQE UCCSD is retained, but the behavior of the ADAPT-VQE infidelities is much more erratic. Secondly, while the ADAPT-VQE ansatz for NaH around the Coulson-Fischer point is mostly the same, but the larger number of variational parameters make it more vulnerable to the optimization inconsistencies discussed above, here the large oscillations are due to ansatze of alternating operator compositions. Because the ADAPT-VQE convergence criterion depends upon a fixed numerical threshold, sometimes the ansatz at a given iteration may already be close to convergence, but still not quite below the gradient norm threshold, and upon the addition of an extra operator, the state may be improved significantly in the scale of the plots seen in this Subsection.

{\color{black}
\subsection{Resource estimation}
\label{ssec:resource}

One of the main motivations behind the present work is to serve as the baseline for following studies focusing on the investigation of the electronic structure of molecules carried out in NISQ devices. With this in mind, it is important to develop some intuition on the resource demands involved in such tasks.

First we analyze the circuit proposed by VQE and ADAPT-VQE to prepare the states whose energies and fidelities were shown in Subsections \ref{ssec:pec}-\ref{ssec:fidelities} in terms of total gate count and circuit depth, plotted in Figure \ref{fig:circuits}.

\begin{figure}[ht!]
    \centering
    \includegraphics[width=\columnwidth]{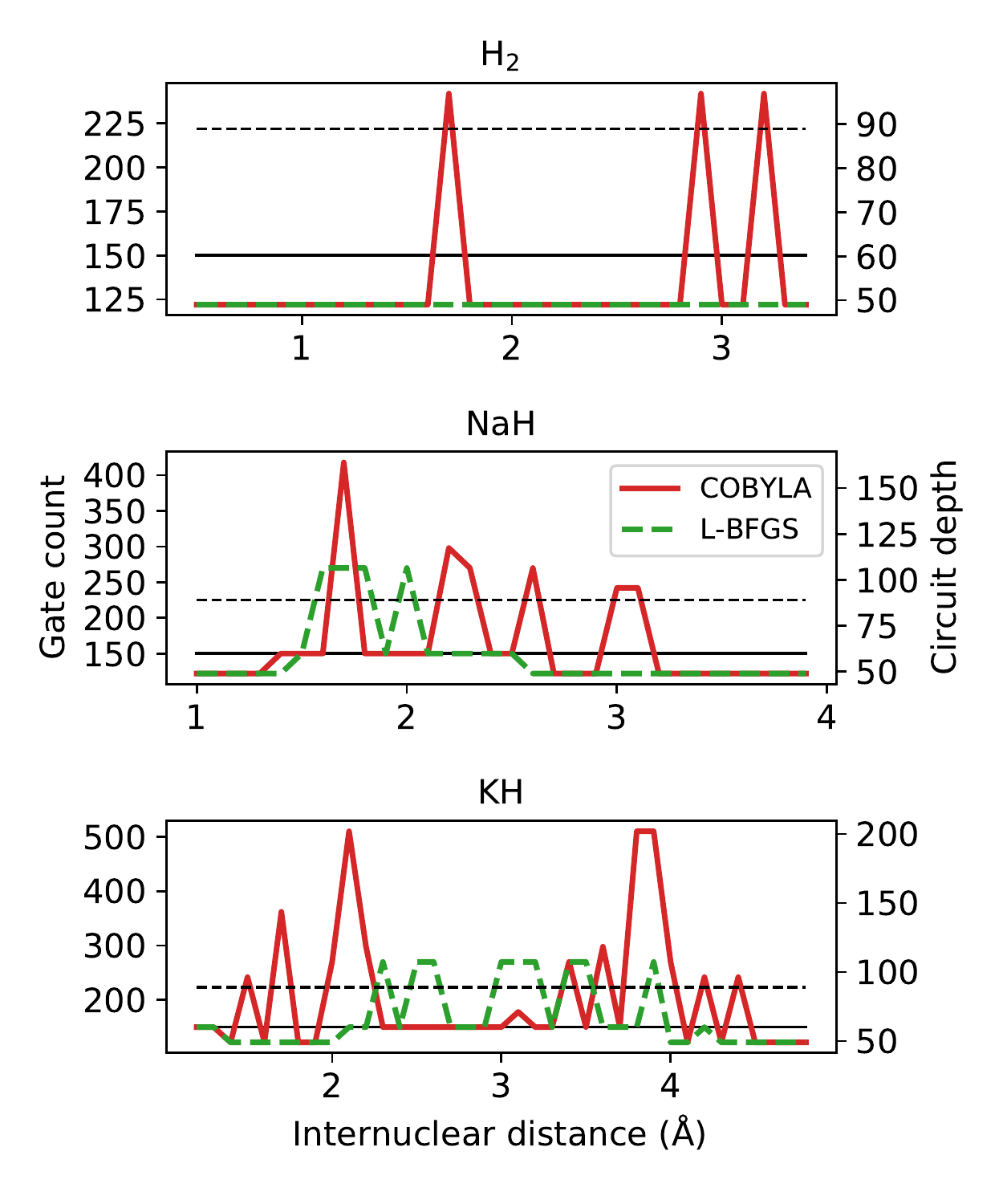}
    \caption{Gate counts (left axis) and circuit depth (right axis) from ADAPT-VQE circuits optimized with the COBYLA and L-BFGS optimizers. The black solid and dashed lines are the gate counts and circuit depths from the VQE ansatz, respectively.}
    \label{fig:circuits}
\end{figure}

Let us first compare the ADAPT-VQE results on the basis of the two optimizers. As we move from H$_2$ to NaH and KH, we see a more intricate picture of how these optimizers impact the final circuit. Qualitatively, L-BFGS has an overall advantage as it provides circuits that are shallower and with fewer gates. While there are a few points along the potential energy scans where the circuits generated based on L-BFGS are not as efficient as those from a COBYLA optimization, the scales of the plots are determined solely by the latter. We noticed that in several points, the simulations with the COBYLA optimizer would produce states with two instances of the same operator adjacent to each other. If the actual minimum value had been achieved in a certain iteration of ADAPT-VQE, the commutator of the same operator in the next iteration would have been zero. Because this procedure is accomplished numerically, the magnitude of this commutator is related to how close the determined minimum is from the actual one. It turns out that the default threshold in relative energy ($10^{-6}$) is found not to be stringent enough, which incurs a commutator whose deviation from the expected zero is non-negligible, resulting in the same operator being added in successive iterations. Another factor that accounts for the displayed circuit figures is the fixed gradient norm threshold in ADAPT-VQE. In some iterations, this quantity is above, but already quite close to the pre-defined $10^{-2}$, and one extra iteration is performed, with only marginal energy improvement. To illustrate this, the ADAPT-VQE simulation for NaH with internuclear separation of 1.8{\AA} converges to ansatz with three operators, with $E = -160.3146751$ Hartree and $||\mathcal{G}|| = 0.001$. Had the ADAPT cycle been stopped in the second iteration, we would have $||\mathcal{G}|| = 0.013$, with $E = -160.3146492$ Hartree, that is, the energy improvement was in the $\mu$Hartree range, yet at the expense of a deeper circuit, which calls for a more flexible operator selection in ADAPT-VQE.

These resource estimation parameters in Figure \ref{fig:circuits} are comparable between the two ansatz strategies. In general terms, the circuits optimized upon L-BFGS are more affordable than the corresponding VQE ones, while using COBYLA tends to yields circuits that are deeper and need to implement more gates. We bring attention to the fact that there is not a one-to-one correspondence between the present analysis and that in the Figures 2c, 2f, and 2i in Ref. \citenum{adapt}. This is because the latter refers to the number of parameters/operators in the ansatz. A circuit with more parameters/operators does not readily translate into a more complex circuit, which depends on the number of qubits in a given operator and the operator locality and placement. This means these results are not at odds with what was previously reported, which were obtained for a distinct set of molecules, but can be seen as complementary.

Another important metric when estimating the necessary resources for implementation and deployment of the simulations discussed here is the number of measurements. To complement the end of the last paragraph, it is important to mention that in this context the rationalization in terms of number of operators increases in relevance. In Figure \ref{fig:measurements} we plot the total number of measurements to achieve the results reported in Subsection \ref{ssec:pec}.

\begin{figure}[ht!]
    \centering
    \includegraphics[width=\columnwidth]{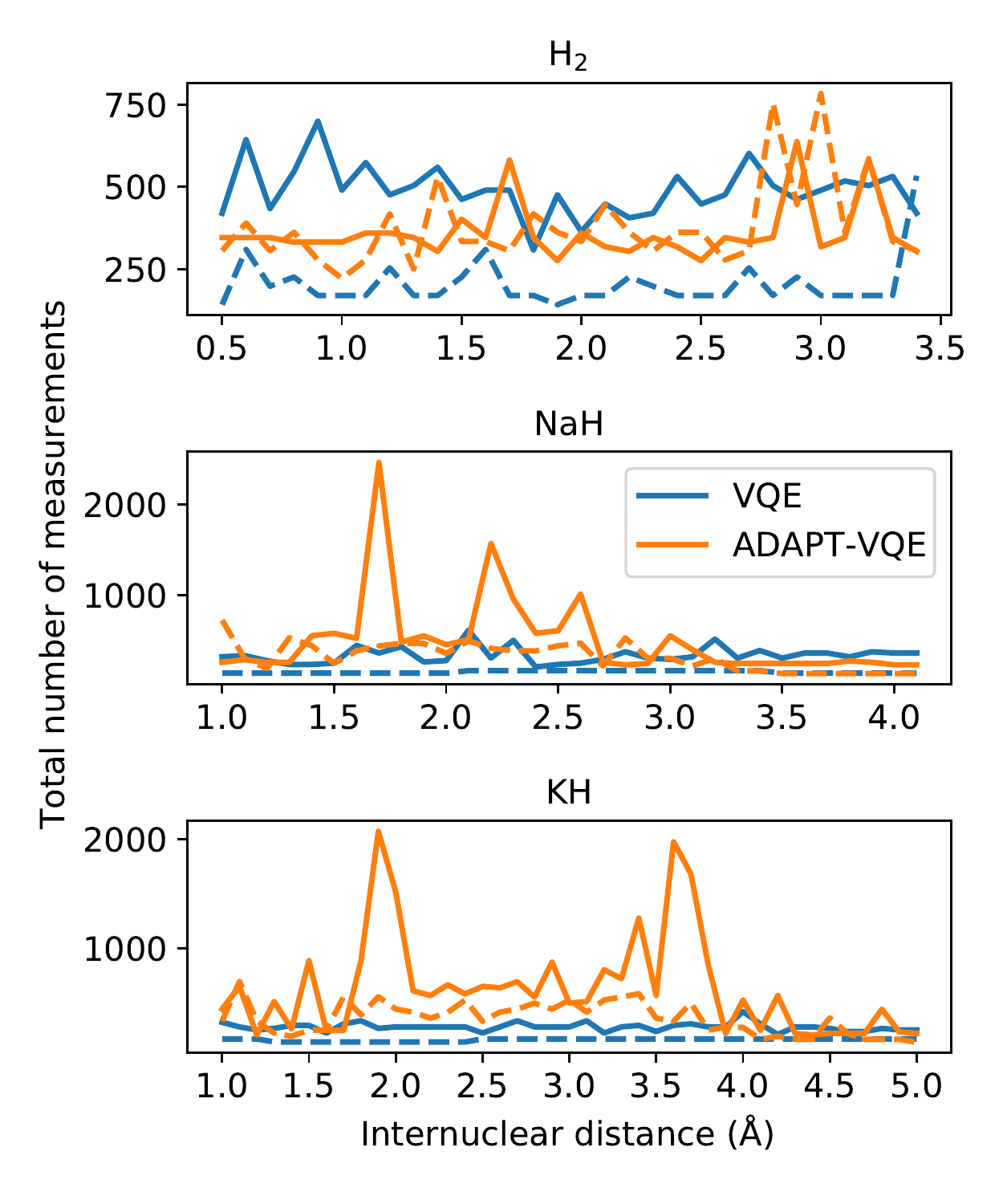}
    \caption{Total number of measurements for final energy evaluation for the VQE and ADAPT-VQE ansatze using the COBYLA (solid line) and L-BFGS (dashed line) optimizers.}
    \label{fig:measurements}
\end{figure}

As pointed out in Ref. \citenum{adapt}, the ansatz put forth by ADAPT-VQE offers a trade-off between circuit depth and number of measurements. We can readily confirm by visual inspection of Figure \ref{fig:measurements} that ADAPT-VQE incurs a much larger number of measurements. These figures account for all measurements involved in computing the commutators in Equation \ref{eq:commutators}, the energy evaluations at each optimization iteration, and the computations necessary to minimize the gradients when L-BFGS is employed. The measurement burden in ADAPT-VQE reported here can be partially alleviated by employing a better parameter initialization, such as starting the VQE optimization at each iteration with the previously optimized parameters and initializing just the newly added parameter at zero. This demand is also expected to be greatly relieved by resorting to a different set of operators, such as those introduced in the qubit-ADAPT-VQE variant,\cite{qubit_adapt} which can still span the underlying Hilbert space, yet with linear growth in the number of qubits. This approach would require much fewer commutator computations at each iteration, but would likely be of noticeable advantage for operator pools larger than those in question here. These results are also contingent upon the choice of optimizer, and there may exist better suited choices than those investigated here. Yet, we do not believe this would dramatically change the overall qualitative picture drawn in Figure \ref{fig:measurements}.

Another key outcome from the analysis of Figure \ref{fig:measurements} is the fact that, even though the gradient computation with L-BFGS requires more measurements per iteration, it is overall much more economical than the gradient-free optimization, represented here by COBYLA. This furthers strengthens the case for gradient-based optimization in VQE, as it not only results in smaller errors/better convergence with respect to the sought ground state, but it is also much less demanding from a resource standpoint.}

\section{Conclusion}
\label{sec:conclusion}

For a broader adoption of adaptive methods for ansatz construction in the realm of quantum chemistry, and perhaps, for many-body methods in general, many aspects still needs to be explored and their underpinnings better understood. This work provides a contribution toward this goal by showing a comprehensive study of potential energy curves of a selection of molecules of the general formula XH (X = H, Na, K). Despite their simplicity, they serve to shed light on some of the mentioned characteristics, and deliver a baseline for feasible studies involving actual quantum hardware.

Even a relatively conservative gradient norm threshold of $10^{-2}$ in ADAPT-VQE is sufficient to provide overall better energetics than corresponding fixed ansatz approach embodied by the ordinary VQE, which is in agreement with the initial ADAPT-VQE proposal. {\color{black}Due to its iterative nature, ADAPT-VQE has an extra layer of tunability which can be controlled via the threshold on $||\mathcal{G}||$. This means that the errors observed with ADAPT-VQE might have been reduced had $||\mathcal{G}||$ been made tighter, which could in turn increase the depth of the circuits, and even having to cope with more necessary measurements than those of UCCSD, as suggested with $||\mathcal{G}|| = 10^{-3}$ in Figure 2i by Grimsley et al.\cite{adapt}} However, upon a simple choice of gradient strategy motivated by the constraints of quantum hardware, we report that ADAPT-VQE is fairly resilient with respect to the employed optimization strategy and that encouraging improvements in performance by adopting a gradient-based approach in the search of the parameter set that minimizes the objective function can be mostly beneficial in the case of VQE. These findings call for a follow-up study on the role of optimizer in conjunction with ADAPT-VQE, extending the analysis to a larger selection of optimizers and gradient strategies.
\par
The ongoing development of VQE methods, including ADAPT-VQE, must also address the noise that is intrinsic to the operations implemented in experimental quantum computers. The above benchmarks of infidelity and energy error place lower bounds on the expected accuracy for VQE methods using noiseless numerical simulation. However, we anticipate that the introduction of noise will substantially affect the accuracy with which the prepared ansatz state approaches the pure state expected from conventional quantum chemistry theory. However, if the state infidelity grows with increasing molecular size, as indicated by our short series of examples, then lower bounds on ansatz accuracy may become a non-trivial contribution to observed errors in experimental measurements. 
\section*{Acknowledgments}
This work was supported by the “Embedding Quantum Computing into Many-body Frameworks for Strongly Correlated Molecular and Materials Systems” project, which is funded by the U.S. Department of Energy (DOE), Office of Science, Office of Basic Energy Sciences, the Division of Chemical Sciences, Geosciences, and Biosciences. This research used resources of the Oak Ridge Leadership Computing Facility, which is a DOE Office of Science User Facilities supported by the Oak Ridge National Laboratory under Contract DE-AC05-00OR22725. This research used resources of the Compute and Data Environment for Science (CADES) at the Oak Ridge National Laboratory, which is supported by the Office of Science of the U.S. Department of Energy under Contract No. DE-AC05-00OR22725. This work was carried out at Oak Ridge National Laboratory, managed by UT-Battelle, LLC for the U.S. Department of Energy under contract DE-AC05-00OR22725.

\bibliographystyle{iopart-num}
\bibliography{paper}

\end{document}